# The Transfer Pricing Problem with Non-Linearities

Svetlana Zverovich

Bristol Business School, University of the West of England, Bristol BS16 1QY, UK

svetlana.zverovich@uwe.ac.uk

ABSTRACT  A number of approaches to solving the well-known transfer pricing problem are known. However, few models satisfactorily resolve the core problem of allowing both the source and receiving divisions to earn a profit on transfers during a period in such a way that sub-optimal output levels are avoided. In 1969, Samuel proposed to use a transfer price schedule instead of just a single transfer price. An essential improvement of Samuels' model was given by Tomkins (1990) in his pragmatic-analytical transfer pricing approach, which is a combination of a single cost-plus transfer price and the pragmatic process of negotiation. This fundamental approach was developed under the assumption that the net average revenue curve for the final product is linear.

In this paper, Tomkins' pragmatic-analytical model is further developed for non-linear net average revenue curves. In particular, typical quadratic functions are considered and corresponding transfer price schedules are determined. A similar technique can be used for the transfer pricing problem with any net average revenue curve.

# 1. INTRODUCTION

Dramatic changes within the business environment in past decades, including decentralisation and globalisation of business, have increased the importance of transfer pricing. Decentralisation implies that a central management cannot monitor and control all the operation parameters of each division and, therefore, it cannot calculate the optimal transfer prices taking into account that each division is usually an autonomous unit. Also, divisional managers have private information, which they may wish to conceal.

However, the transfer pricing mechanism applied by an organisation can have a critical impact on its performance. This mechanism should motivate divisional managers to make optimal economic decisions without undermining divisional autonomy, and provide a reasonable measure for evaluating the managerial and economic performance of the source and receiving divisions, while being acceptable for taxation purposes.

A number of approaches to resolving the transfer pricing conflicts are known. For example, goods can be transferred between the divisions at a variable cost with profit to the supplier provided by a periodic charge. However, this two-part transfer price system has a serious disadvantage because "the source division has no incentive to seek the optimal production level as it earns no profit on transactions made *during* the period" (Tomkins, 1990, p. 202). Another approach is based on sharing the group profit earned on the transferred goods, but its shortcomings include problems with the book-keeping and the economic meaning of the profit. Also, a more advanced dual-rate transfer pricing system has been proposed. The problems with this approach concern the artificial nature of dual-rate transfer prices, which can lead to confusion. As a result, they are not widely used in practice. Many decentralized organizations use simple negotiation or negotiated optimal two-part tariffs to determine transfer prices. Although the efficiency of the latter is significantly higher than the efficiency of simple direct negotiation (Lantz, 2009), the problem with this approach is that the outcome may depend on the managers' negotiation skills and their unequal bargaining power. Recently, an interesting descriptive unconstrained model for transfer pricing in multinational supply chains has been proposed, which takes into account many factors (Villegas and Ouenniche, 2008). However, the authors have not provided any specific solution procedure to the model as "such a procedure would require specific assumptions about the cost and the revenue functions taken into account in the paper" (Villegas and Ouenniche, 2008, p. 846). Transfer pricing has been also studied by Gjerdrum et al. (2002), Lakhal (2006), Lakhal et al. (2005), Li (1997), Pfeiffer (1999), and Vidal and Goetschalckx (2001).

Thus, the well-known transfer pricing problem has been studied over a long period of time. However, few models satisfactorily resolve the core problem of allowing both the source and receiving divisions to earn a profit on transfers during a period in such a way that sub-optimal output levels are avoided. Also, there is no explanation why "the theoretical recommendations differ from the practice of transfer pricing, with the possible exception of Tomkins' (1991) pragmatic-analytical perspective" (McAulay and Tomkins, 1992, p. 116).

In 1969, Samuel proposed the use of a transfer price schedule instead of simply a single transfer price. Tomkins (1990, p. 203) criticises this as follows:

> The trouble with Samuels' proposal is that one needs to know the optimal level of production and the amount to be transferred before one can see what point the pricing schedule needs to pass through… Also, while Samuels devised this basic concept some time ago, there is no evidence of its wide adoption in industry. One can only assume that it is seen by companies as too complex.

An essential improvement of Samuels' model was given by Tomkins (1990) in his pragmatic-analytical transfer pricing approach, which is a combination of a single cost-plus transfer price and the pragmatic process of negotiation. In an attempt to understand why practitioners use full costing he has proved that full costing can provide optimum results, where full-cost transfer prices are used over the majority of output transferred and are negotiated over a small proportion of transfers.

Tomkins' fundamental approach was developed under the assumption that the net average revenue curve for the final product is linear. However, in practice this is not the case:

> The demand curve is normally drawn in textbooks as a straight line suggesting a linear relationship between price and demand but in reality, the demand curve will be non-linear! No business has a perfect idea of what the demand curve for a particular product looks like, they use real-time evidence from markets to estimate the demand conditions and their accumulated experience of market conditions gives them an advantage in constructing demand-price relationships. (Riley, 2006, p. 25)

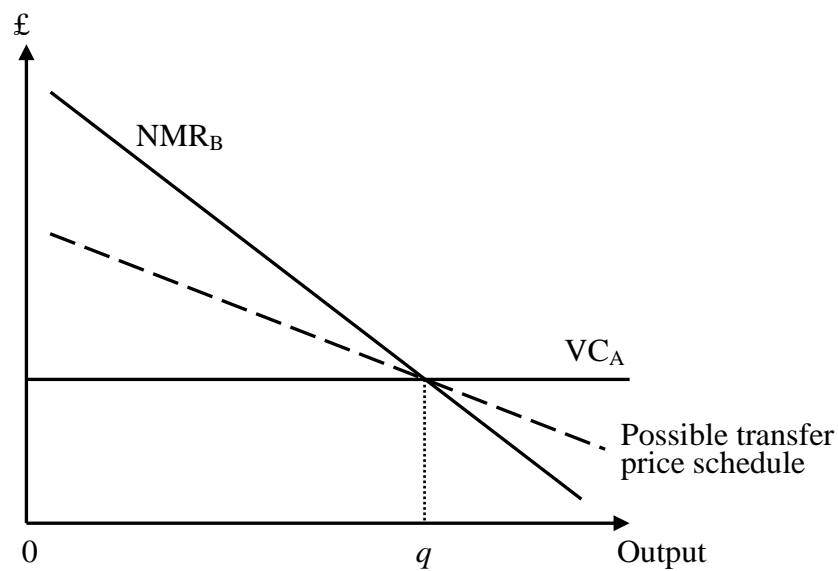

**Figure 1.** Samuels' transfer pricing policy.

Tomkins (1990) discussed a non-linear net average revenue curve and considered a convex quadratic curve as a natural generalisation of a linear function. However, no solution for quadratic functions was given. In our view, a convex quadratic curve is indeed the most typical non-linear net average revenue curve, but it is unclear what is a shape of a typical concave net average revenue curve, which was also mentioned by Tomkins (1990). In this paper, we will consider convex quadratic functions and an example of a concave function, and determine the corresponding transfer price schedules. Moreover, in Section 5 we explain how a similar technique can be used to solve the transfer pricing problem with any specified net average revenue curve.

## 2. TOMKINS' PRAGMATIC-ANALYTICAL MODEL

We assume that Division A, a source division, produces goods and transfers them to Division B, i.e. there is no intermediate good market, and Division B, after further processing, sells the goods to an external market. It is also assumed that Division A is a profit centre as it produces other goods that are not transferred to Division B, and for those goods there is an intermediate external market. Let $q$ be the optimal level of output to be produced by A and transferred to B, $VC_A$ stand for A's variable cost, and $NMR_B$ denote B's net marginal revenue.

Suppose that one wants both Division A and Division B to earn a profit on transfers in order to motivate them to optimally produce and transfer goods. To solve the problem, Samuels (1969) proposes a transfer price schedule as indicated in Figure 1. It is well known that this schedule encourages both A and B to use the optimal level $q$, which maximises overall profits.

Tomkins (1990) points out a number of disadvantages of Samuels' model. The most serious trouble is that "one needs to know the optimal level of production and the amount to be transferred before one can see what point the pricing schedule needs to pass through" (Tomkins, 1990, p. 203). Another problem is that this model has not been widely adopted in industry – perhaps it is too complex to use, and therefore a simpler approach is needed.

An improvement of Samuels' model was given by Tomkins (1990) in his pragmatic-analytical transfer pricing approach, which is a combination of a single transfer price and the pragmatic process of negotiation as illustrated in Figure 2. The basic idea is that a company fixes a transfer price $t$ by a cost-plus method provided that the transfer price is limited to an output level $f$ such that the following condition is satisfied (Tomkins, 1990, p. 207):

> *Constraint 1*
> The transfer price line projected horizontally to the right must not cut through B's net marginal revenue schedule if sub-optimal output levels are to be avoided.

The transfer price $t$ must be fixed in such a way that A's fixed costs are covered and, moreover, a 'reasonable' profit is provided to A, see Figure 2. The area of negotiation (between $f$ and $q$) is where the transfer price $t$ is no longer applicable, i.e. A and B should negotiate further transfers. The scale of negotiation must satisfy the following condition (Tomkins, 1990, p. 207):

> *Constraint 2*
> The cost-plus transfer price (*t*) must be applicable to a sizeable proportion of the optimal amount of output (*f*) to be transferred in order to limit the scale of negotiation needed.

Thus, the objective is to maximise the proportion of optimal output, denoted by $x$, for which the cost-plus transfer price can be applied, subject to Constraint 1 and the condition that A's fixed costs are covered and A's profit is provided. Let $c$ stand for source division's target contribution as a proportion of maximum group contribution.

Tomkins (1990) assumed the following:

- Division A transfers goods to Division B, i.e. there is no intermediate good market.
- All of A's and B's costs are fixed.
- The net average revenue curve for the final product ($NAR_B$) is linear.

Note that the second assumption is not realistic and it was only made for simplicity of calculations. However, Tomkins (1990) gave no adjustment of the optimal transfer price schedule when this assumption does not hold, i.e. variable costs are not equal to zero. We will make the same assumptions and then show how to adjust the optimal schedule to take into account variable costs.

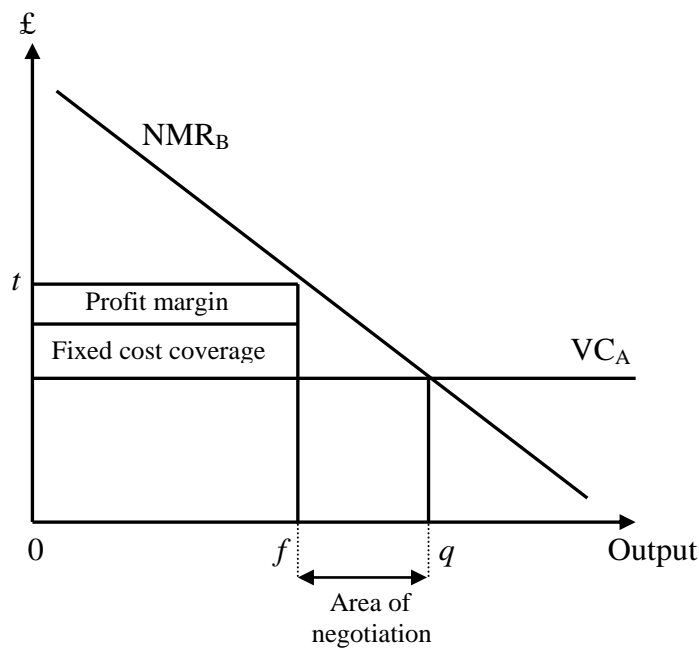

**Figure 2.** Tomkins' pragmatic-analytical transfer pricing approach.

Under these assumptions, Tomkins (1990) proved an interesting result that, for a given $c$, the value of $x$ can be found by solving the following quadratic equation:

$$2x(1-x) - c = 0.$$

In fact, $x$ is the upper root of the above equation, i.e.

$$x = 0.5 + \sqrt{0.25 - 0.5c}.$$

Note that such a root exists only if 0≤*c*≤0.5. Now, the corresponding transfer price can be calculated as follows:

$$t = 2p(1-x),$$

where *p* is the net average revenue corresponding to the output *q*. Tomkins (1990) summarises his result in Table 1:

| Source division's target contribution as a proportion of maximum group contribution (*c*) | Maximum proportion of optimal output for which the cost-plus transfer price can be applied (*x*) | Proportion of group contribution over which negotiation required (*n*) |
|---|---|---|
| 0.50 | 0.50 | 0.250 |
| 0.48 | 0.60 | 0.160 |
| 0.46 | 0.64 | 0.129 |
| 0.42 | 0.70 | 0.090 |
| 0.38 | 0.74 | 0.065 |
| 0.32 | 0.80 | 0.040 |
| 0.26 | 0.85 | 0.024 |
| 0.18 | 0.90 | 0.010 |
| 0.09 | 0.95 | 0.002 |
| 0.00 | 1.00 | 0.000 |

**Table 1.** The likely practical relevance of the pragmatic-analytical approach.

For example, if A's target contribution is set at 18% of the total maximum group contribution, then the corresponding cost-plus transfer price can be applied to 90% of the optimal output to recover its target contribution. As can be seen in Table 1, the proportion of group contribution over which negotiation is required is only 1%, so the negotiation makes practically no difference. In contrast, if A's target contribution is close to 50% of the total maximum group contribution, then the proportion of group contribution over which negotiation is required becomes substantial. For instance, if *c* is equal to 50%, then negotiation is required over 25% of the group contribution.

For convenience, let us summarize the notation introduced in this section:

| | |
|---|---|
| *q* | The optimal level of output to be produced by Division A and transferred to B. |
| $VC_A$ | Division A's variable cost. |
| $NMR_B$ | Division B's net marginal revenue. |
| *p* | B's net average revenue corresponding to the output *q*. |
| *pq* | The maximum contribution, which the entire group can earn. |
| *c* | A's target contribution as a proportion of maximum group contribution. |
| *t* | The transfer price. |
| *f* | B's chosen output level corresponding to the transfer price *t*. |
| *x* | Maximum proportion of optimal output for which the cost-plus transfer price can be applied. |
| *n* | Proportion of group contribution over which negotiation required. |

**Table 2.** The basic parameters.

## 3. QUADRATIC NAR$_B$ CURVE

In this section, we assume that there is no intermediate good market, all A's and B's costs are fixed, but that B's NAR curve for the final product is quadratic and convex from above, i.e.

$$\text{NAR}_B(f) = a - bf^2,$$

where $a, b > 0$. The total revenue, denoted by TR, is

$$\text{TR}(f) = af - bf^3,$$

and therefore the NMR$_B$ schedule is

$$\text{NMR}_B(f) = \frac{\partial \text{TR}}{\partial f} = a - 3bf^2.$$

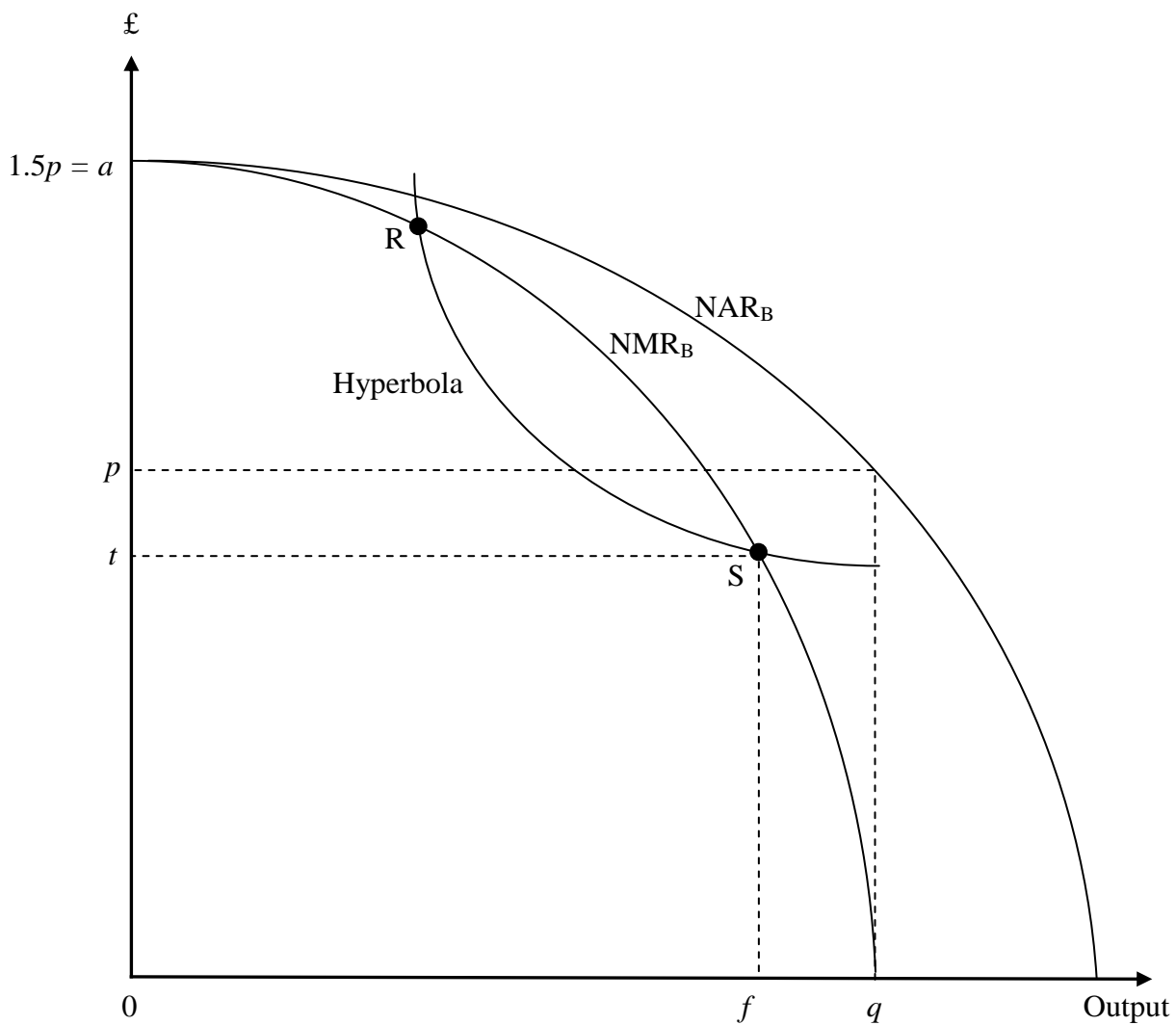

**Figure 3.** An illustration of a quadratic NAR$_B$ curve.

We remind that $x = f/q$ denotes the proportion of optimal output for which the transfer price $t$ is applied. For a given proportion $c$, there are multiple $t \times f$ rectangles with an area equal to $c$, i.e. $tf = c(pq)$. Therefore,

$$t = \frac{cpq}{f} = \frac{cp}{f/q} = \frac{cp}{x}.$$

Thus, the top right corners of such $t \times f$ rectangles are specified by the hyperbola RS (see Figure 3) with the following equation:

$$t = \frac{cp}{x},$$

where $c$ and $p$ are fixed numbers, and $t$ and $x$ are variable. To promote efficiency an aim is to maximise $x$, so we need to find the coordinates of the point S. In order to find them, we need to equate $cp/x$ to $a - 3bf^2$:

$$\frac{cp}{x} = a - 3bf^2.$$

Substituting $xq$ for $f$, we obtain

$$3bq^2 x^3 - ax + cp = 0. \tag{1}$$

By definition, $p = a - bq^2$, and, by rearranging, we have

$$bq^2 = a - p.$$

The NMR$_B$ schedule gives $0 = a - 3bq^2$. Hence $0 = a - 3(a - p)$ or

$$a = 1.5p. \tag{2}$$

Therefore,

$$bq^2 = a - p = 0.5p. \tag{3}$$

Thus, equation (1) can be rewritten as follows:

$$1.5px^3 - 1.5px + cp = 0$$

or

$$x^3 - x + \frac{2}{3}c = 0. \tag{4}$$

We need to find the upper root out of two roots between 0 and 1. The function $f(x) = x^3 - x + 2c/3$ has its minimum between 0 and 1 at the point $x = \sqrt{1/3}$. Hence, the two roots between 0 and 1 only exist if

$$(1/3)^{1.5} - (1/3)^{0.5} + 2c/3 < 0$$

or

$$c < \sqrt{1/3} \approx 0.577.$$

In what follows, we assume that $c < \sqrt{1/3}$. Now we apply a standard technique to solve equation (4). The discriminant of the above cubic equation (2) is $D = Q^3 + R^2$, where $Q = -\dfrac{1}{3}$ and $R = -\dfrac{c}{3}$. Therefore,

$$D = \frac{1}{9}\left(c^2 - \frac{1}{3}\right).$$

Since $c < \sqrt{1/3}$, the discriminant $D$ is less than 0, and hence all three roots are real and unequal.

If we put

$$\theta = \arccos\left(\frac{R}{\sqrt{-Q^3}}\right) = \arccos\left(-c\sqrt{3}\right),$$

then the real roots are:

$$x_1 = 2\sqrt{-Q}\cos\left(\frac{\theta}{3}\right) = \frac{2}{\sqrt{3}}\cos\left(\frac{\theta}{3}\right),$$

$$x_2 = 2\sqrt{-Q}\cos\left(\frac{\theta + 2\pi}{3}\right),$$

$$x_3 = 2\sqrt{-Q}\cos\left(\frac{\theta + 4\pi}{3}\right).$$

It is easy to check that $x_2$ is a negative root and $0 \leq x_3 < x_1 \leq 1$, i.e. $x_1$ is the required root.

Let $n$ denote the proportion of group contribution over which negotiation is required. Taking into account (2) and (3), it is not difficult to see that

$$npq = \int_f^q (a - 3bf^2)\,df$$

$$= \left[(af - bf^3)\right]_{xq}^{q}$$

$$= q(a - bq^2) - xq(a - bx^2q^2)$$

$$= qp - xq(1.5p - x^2 0.5p)$$

$$= qp(1 - 1.5x + 0.5x^3).$$

Therefore,

$$n = 1 - 1.5x + 0.5x^3.$$

Some combinations of the parameters *c*, *x* and *n* are summarised in Table 3.

| Source division's target contribution as a proportion of maximum group contribution ($c$) | Maximum proportion of optimal output for which the cost-plus transfer price can be applied ($x$) | Proportion of group contribution over which negotiation required ($n$) |
|---|---|---|
| 0.57 | 0.630 | 0.180 |
| 0.55 | 0.677 | 0.140 |
| 0.50 | 0.742 | 0.091 |
| 0.45 | 0.786 | 0.064 |
| 0.40 | 0.822 | 0.045 |
| 0.35 | 0.852 | 0.031 |
| 0.30 | 0.879 | 0.021 |
| 0.25 | 0.903 | 0.014 |
| 0.20 | 0.925 | 0.008 |
| 0.15 | 0.946 | 0.004 |
| 0.10 | 0.965 | 0.002 |
| 0.05 | 0.983 | 0.000 |
| 0.00 | 1.000 | 0.000 |

**Table 3.** The practical relevance of the quadratic model.

### 3.1. Transfer Price Schedule

For a given $c$, the first step is to calculate $x$, the proportion of optimal output for which the transfer price $t$ is applied:

$$x = \frac{2}{\sqrt{3}} \cos\left(\frac{1}{3} \arccos\left(-c\sqrt{3}\right)\right). \quad (5)$$

The transfer price $t$ can be found using this formula:

$$t = \frac{cp}{x}. \quad (6)$$

The formulae (5) and (6) provide the necessary transfer price schedule. Divisions A and B should negotiate further transfers over the following proportion of group contribution:

$$1 - 1.5x + 0.5x^3.$$

### 3.2. Transfer Price Schedule with Non-Zero VC$_A$ and an Example

If A's variable cost is a non-zero constant, then the precise meaning of $c$ becomes slightly different: it is A's target contribution less A's total variable costs as a proportion of maximum group contribution less A's total variable costs. Now let $c_{\text{real}}$ denote A's target contribution as a proportion of maximum group contribution. We have

$$c_{\text{real}}\, qp = cq(p - VC_A) + qVC_A.$$

Therefore,

$$c = \frac{pc_{\text{real}} - VC_A}{p - VC_A}. \tag{7}$$

Next, the proportion of optimal output $x$, for which the transfer price $t$ is applied, can be calculated by formula (5) or taken from Table 3. The transfer price $t$ can be found using the following formula:

$$t = \frac{c(p - VC_A)}{x} + VC_A. \tag{8}$$

The formula for the proportion of group contribution over which negotiation is required is as follows:

$$n(1 - VC_A/p).$$

For example, suppose that A's target contribution is 40%, i.e. $c_{\text{real}}$ is set at 0.4, and let $p$=£100 and $VC_A$=£20. Then formula (7) yields $c$=0.25. From Table 3 or by formula (5), we obtain $x$=0.903. Thus, using (8), we have $t$ = £42.15. The negotiation is required over $1.4(1-0.2) = 1.1\%$.

Note that the above formulae can be easily implemented in Excel. For example, assuming that $c$ is kept in cell A1, formula (5) can be written in Excel as follows:

=(2/SQRT(3))*COS(ACOS(-A1*SQRT(3))/3)

## 4. EXPONENTIAL NAR$_B$ CURVE

Although in reality the NAR curve is rather convex and quadratic, we would like to consider an example illustrating that the above technique also works for non-quadratic non-convex NAR curves.

Let us make the same assumptions, except that the NAR curve for the final product is now exponential and concave from above:

$$\text{NAR}_B(f) = ae^{-f/b},$$

where $a>0$ and $b>0$. It follows that the total revenue is

$$\text{TR}(f) = fae^{-f/b}.$$

Therefore,

$$\text{NMR}_B(f) = \frac{\partial \text{TR}}{\partial f} = ae^{-f/b}(1 - f/b).$$

Since NMR$_B(q)$=0, we obtain $1 - q/b = 0$, i.e. $b = q$. Also, NAR$_B(q) = p$. Hence

$$ae^{-q/q} = p, \text{ i.e.}$$

$$a = ep.$$

Thus,
$$\text{NAR}_B(f) = pe^{1-f/q}$$

and
$$\text{NMR}_B(f) = pe^{1-f/q}(1 - f/q).$$

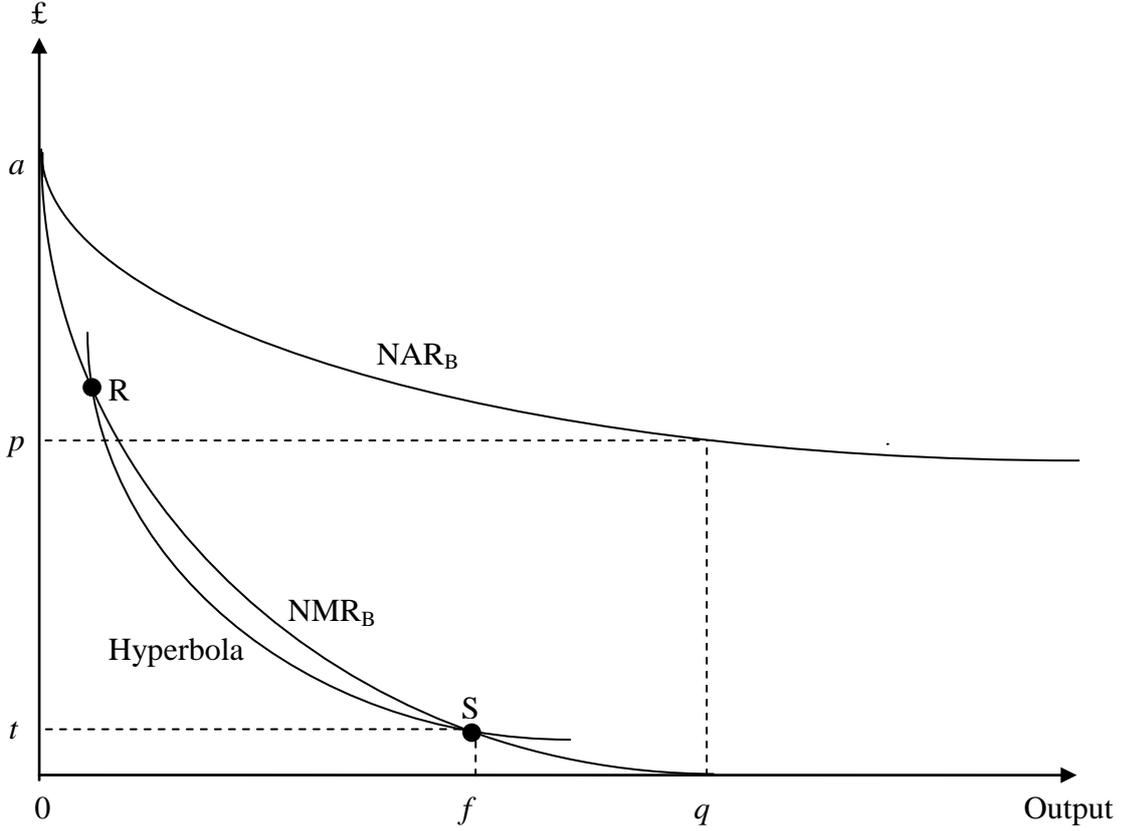

**Figure 4.** An illustration of an exponential NAR$_B$ curve.

Similar to Section 3, we need to find the coordinates of the point S. We have

$$\frac{cp}{x} = pe^{1-x}(1-x)$$

or
$$x(1-x)e^{1-x} = c. \tag{9}$$

The proportion of group contribution over which negotiation is required can be found as follows:

$$npq = \int_{xq}^{q} pe^{1-f/q}(1 - f/q)df$$

$$= \left[fpe^{1-f/q}\right]_{xq}^{q}$$

Therefore,
$$= qp(1 - xe^{1-x}).$$

$$n = 1 - xe^{1-x}. \tag{10}$$

For convenience, a number of combinations of the parameters $c$, $x$ and $n$ are shown in Table 4:

| Source division's target contribution as a proportion of maximum group contribution ($c$) | Maximum proportion of optimal output for which the cost-plus transfer price can be applied ($x$) | Proportion of group contribution over which negotiation required ($n$) |
|---|---|---|
| 0.43 | 0.446 | 0.224 |
| 0.42 | 0.480 | 0.193 |
| 0.40 | 0.527 | 0.154 |
| 0.38 | 0.565 | 0.127 |
| 0.36 | 0.597 | 0.107 |
| 0.34 | 0.626 | 0.090 |
| 0.32 | 0.654 | 0.076 |
| 0.30 | 0.680 | 0.064 |
| 0.28 | 0.704 | 0.053 |
| 0.26 | 0.728 | 0.044 |
| 0.24 | 0.751 | 0.037 |
| 0.22 | 0.773 | 0.030 |
| 0.20 | 0.795 | 0.024 |
| 0.18 | 0.817 | 0.019 |
| 0.16 | 0.838 | 0.015 |
| 0.14 | 0.858 | 0.011 |
| 0.12 | 0.879 | 0.008 |
| 0.10 | 0.899 | 0.005 |
| 0.08 | 0.920 | 0.003 |
| 0.06 | 0.940 | 0.002 |
| 0.04 | 0.960 | 0.001 |
| 0.02 | 0.980 | 0.000 |
| 0.00 | 1.000 | 0.000 |

**Table 4.** The practical relevance of the exponential model.

### 4.1. Transfer Price Schedule

Thus, for a given A's target contribution $c$, the corresponding maximum proportion $x$ of the optimal output, for which the transfer price $t$ can be applied, is found from Table 4 or by solving equation (9). Then, the transfer price $t$ can be found using this formula:

$$t = \frac{cp}{x}.$$

Divisions A and B should negotiate further transfers over the proportion of group contribution, which can be found by formula (10). Similar to Section 3.2, the above formulae can easily be adapted for the case when $VC_A$ is not equal to zero.

## 5. CONCLUDING REMARKS

In this paper, Tomkins' pragmatic-analytical model was further developed for non-linear net average revenue curves. In our view, the most typical non-linear net average revenue curves are convex quadratic functions. These have been considered and corresponding transfer price schedules have been determined. We have also considered an exponential NAR curve to illustrate that the above technique works for non-quadratic non-convex curves. Note that both the quadratic and exponential NAR curves depend on two coefficients *a* and *b*, but the corresponding transfer price schedules described in Sections 3.1, 3.2 and 4.1 are not dependent on those coefficients. This implies that in practice we only need to decide whether the NAR curve is quadratic (exponential), and there is no need to determine the coefficients *a*, *b* and the exact formula for such a function.

A similar technique can be used for any NAR curve. Indeed, if we know $k+1$ points of the curve, then it can be approximated by the interpolation polynomial in the Lagrange form of degree *k*. Therefore, we can find the $NMR_B$ schedule and then, using the above hyperbola RS, construct a polynomial of degree $k+1$ of the variable *x*. For $k=2$, such a polynomial is similar to the left-hand side of equation (1). We remind that the aim is to find the largest root of this polynomial, which does not exceed 1. Of course, for $k=2$, the cubic polynomial equation (1) can be simplified to (4) and then solved analytically as shown in Section 3. However, in general we cannot find this root analytically, so it must be determined computationally using one of the known methods.

Thus, the fundamental approach of Tomkins can be applied to the transfer pricing problem with any net average revenue curve.